\documentclass[aps,twocolumn,showpacs]{revtex4}
\usepackage{graphics}
\usepackage{latexsym}
\begin{document}
\tighten
\bibliographystyle{apsrev}
\def\half{{1\over 2}}
\def \D {\mbox{D}}
\def\curl {\mbox{curl}\,}
\def \ep {\varepsilon}
\def \lleq {\lower0.9ex\hbox{ $\buildrel &lt; \over \sim$} ~}
\def \ggeq {\lower0.9ex\hbox{ $\buildrel &gt; \over \sim$} ~}
\def\beq{\begin{equation}}
\def\eeq{\end{equation}}
\def\beqa{\begin{eqnarray}}
\def\eeqa{\end{eqnarray}}
\def \apl {ApJ, }
\def \aps {ApJS, }
\def \pd {Phys. Rev. D, }
\def \prl {Phys. Rev. Lett., }
\def \pl {Phys. Lett., }
\def \np {Nucl. Phys., }
\def \l {\Lambda}

\def\D{{\cal D}}
\def\R{{\cal R}}
\def\G{{\cal G}}
\def\H{{\cal H}}
\def\g{\,^{(5)\!}g}
\def\m{{(\!m\!)}}
\def\vu{^{(5)\!}}
\newcommand{\be}{\begin{equation}}
\newcommand{\ee}{\end{equation}}
\newcommand{\bea}{\begin{eqnarray}}
\newcommand{\eea}{\end{eqnarray}}
\title{Aspects of Scalar Field Dynamics in Gauss-Bonnet Brane Worlds}
\pacs{98.80.Cq,~98.80.Hw,~04.50.+h}  
\author{M. Sami}
\altaffiliation[On leave from:]{ Department of Physics, Jamia Millia, New Delhi-110025}
\email{sami@iucaa.ernet.in}
\affiliation{IUCAA, Post Bag 4, Ganeshkhind,\\
 Pune 411 007, India.}
\author{N. Savchenko}
% \affiliation{Sternberg Astronomical Institute, Universitetski Prospect$-$13,\\
% Moscow 119899, Russia}
 \email{savchenko@xray.sai.msu.ru}
\author{A. Toporensky}
 \email{lesha@sai.msu.ru}   
 \affiliation{Sternberg Astronomical Institute, Universitetski Prospect$-$13,\\
 Moscow 119899, Russia}

\begin{abstract}
The Einstein-Gauss-Bonnet equations projected from the bulk to brane lead to a complicated Friedmann equation
which simplifies to $H^2 \sim \rho^q$ in the asymptotic regimes. The Randall-Sundrum (RS) scenario
corresponds to $q=2$ whereas $q=2/3$ $\&$ $q=1$ give rise to high energy Gauss-Bonnet (GB) regime and the standard GR
respectively. Amazingly, while evolving from RS regime to high energy GB limit, one passes through a
GR like region which has important implications for brane world inflation. For tachyon GB inflation with
 potentials $V(\phi) \sim \phi^p$ investigated in this paper, the scalar to tensor ratio of perturbations $R$
is maximum around the RS region and is generally suppressed in the high energy regime for the positive values of $p$. 
The ratio is very low for  $p>0$
at all energy scales relative to GB inflation with ordinary scalar field.
The models based upon tachyon inflation with polynomial type of potentials with generic positive values of $p$
turn out to be in the $1 \sigma$ observational contour bound at all energy scales varying from GR to high energy
GB limit.
The spectral index $n_S$ improves for the lower values of $p$ and approaches its scale invariant limit
for $p=-2$ in the high energy GB regime. The ratio $R$ also remains small for large negative values of $p$, however,
difference arises for models close to scale invariance limit. In this case,
the tensor to scale ratio is large in the GB regime whereas it is suppressed in the intermediate region between RS and GB. 
Within the frame work of patch cosmologies governed by $H^2 \sim \rho^q$, the behavior of ordinary scalar field
near cosmological singularity and the nature of scaling solutions are distinguished for the values of 
$q \, < 1$ and $q\, > 1$. The tachyon dynamics, on the other hand, exhibits stable scaling solutions $\forall q$ if
the adiabatic index of barotropic fluid $\gamma <1$.
\end{abstract}

\maketitle
\section{INTRODUCTION}
Being inspired by D-brane ideology in string theory, the brane world scenario {\it a la} Randall-Sundrum (RS)\cite{randall,h}
envisages that our four dimensional space time (brane) is embedded in the 5-dimensional bulk. To be in line
with string theory, it is assumed that all the standard model degrees of freedom reside on the
brane where as gravity can propagate into bulk. In adherence to Newtonian gravity in the
low energy limit, the bulk is assumed to be anti de-Sitter allowing gravity to be localized near
the brane dynamically and thereby leading to Newton's law with small corrections at large distances. The 
space time dynamics projected from the bulk to brane leads to the modified Einstein equations on the brane. The resulting Hubble equation on the FRW brane ,among other things, contains high energy
corrections which have important implications for early universe physics. In particular,
the prospects of inflation are enhanced in brane world cosmology. In the case of standard FRW , the
steep potentials can not support inflation and bouncing solutions. The presence of the quadratic density term (high energy corrections) in the Friedman equation on the brane changes the
expansion dynamics at early epochs \cite{cline}(see Ref\cite{roy rev} for details on the dynamics of brane worlds). 
Consequently, the field experiences greater damping and rolls down its potential slower than it would during the conventional inflation.
 Thus, inflation in the brane world scenario can successfully occur for very steep potentials\cite{basset,liddle}. The brane assisted inflation
allows to build successful models of quintessential inflation\cite{unified}. However, the recent WMAP observations and large scale galaxy clustering studies
severely constraint  the steep brane world inflation. For instance, the inflation driven by steep exponential potential
in RS scenario is excluded by observation for the number of e-folds as large as $70$\cite{shinji}. It was recently shown that Gauss-Bonnet (GB) correction in the bulk could rescue these
models\cite{ssr}.\\ 
There is a sound theoretical reason to include the higher curvature terms in Einstein-Hilbert action\cite{GB1,GB2}. These terms arise perturbatively as 
next to leading order correction in effective
string theory action. The Gauss-Bonnet combination
is special in five dimensions as it is a unique invariant which leads to field equations of second order linear in the highest derivative thereby ensuring
a unique solution\footnote{The GB term can as well be motivated purely on classical considerations. It arises naturally as
higher order iteration of the self interaction of gravitational field which retains the
quasi-linear second order character of
the field equation. The physical realization of this iteration naturally requires a 5-dimensional
space time\cite{naresh}.}.\\
The Einstein-Gauss-Bonnet equations projected on to the brane lead to a complicated Hubble equation 
in general\cite{D,lidsey,T,DR} (see also Ref\cite{g}) . Interestingly, it
reduces to a very simple equation $H^2 \sim \rho^q$ with $q=1,2,2/3$ in limiting cases corresponding to GR, RS and
GB regimes respectively.
In the high energy GB regime, this allows to push the spectral index $n_S$ very close to one for exponential potential in case of ordinary scalar field \cite{lidsey}. The tachyonic inflation is recently studied in patch cosmologies
in view of observational constraints\cite{gs}. The patches corresponding to GR, RS and GB naturally arise in the dynamical
history described by the exact effective Hubble equation on the brane in presence of the GB term in the bulk.
It is really interesting to carry out the detailed investigations of tachyon field inflation in the full GB dynamics
which gives rise to the mentioned patches at relevant energy scales. It is also important to investigate the behavior
of scalar field near singularity and look for the scaling solutions in the patch cosmologies.\\
In this paper we study different aspects of scalar field dynamics in brane worlds with GB term in the bulk. In section II, we review
the basic concepts of GB brane world cosmology. In section III, we investigate the tachyon  inflation in the GB background with 
polynomial type of potentials which corresponds to  an exponential potential in a special case. This section includes
the detailed description of tachyon field inflation at all the energy scales from GR to high energy GB regime.\\
Section IV
is devoted to the study of non-inflationary dynamics of ordinary scalar field in the background governed by the Friedmann equation
 $H^2 \sim \rho^q$. This section contains the description of asymptotic behavior of scalar field near singularity
and the existence of scaling solutions in the  background cosmology under consideration.

\section{Gauss-Bonnet Brane Worlds}
The Einstein-Gauss-Bonnet action for five dimensional bulk containing a 4D brane is
\begin{eqnarray}
S &=& \frac{1}{2 \kappa_5^2}\int d^5x\sqrt{-g}\big\lbrace {\cal R} - 2\Lambda_5
+ \alpha \lbrack {\cal R}^2 - 4{\cal R}_{AB}{\cal R}^{AB} \nonumber\\
&+& {\cal R}_{ABCD} {\cal R}^{ABCD}\rbrack \big \rbrace
+ \int d^4x\sqrt{-h} ({\cal L}_m - \lambda)~,
\end{eqnarray}
${\cal R}$ refers to the Ricci scalar in the bulk metric $g_{AB}$ and $h_{AB}$ is the induced metric on the brane
; $\alpha$ has dimensions of ({\em length})$^2$ and
is the Gauss-Bonnet coupling, while $\lambda$ is the brane tension
 and $\Lambda_5\,(<0)$ is the bulk
cosmological constant. The constant $\kappa_5$ contains the 5D fundamental energy scale ($\kappa_5^2=M_5^{-3}$).\par
A Friedman-Robertson-Walker (FRW) brane in an AdS$_5$ bulk is a
solution to the field and junction equations~\cite{D}. The
modified Friedman equation on the (spatially flat) brane may be written as~\cite{D,T}
 \begin{eqnarray}
H^2 &=& {1\over
4\alpha}\left[(1-4\alpha\mu^2)\cosh\left({2\chi\over3}
\right)-1\right]\,,\label{mfe}\\
\label{chi} \kappa_5^2(\rho+\lambda) &=&
\left[{{2(1-4\alpha\mu^2)^3} \over {\alpha} }\right]^{1/2}
\sinh\chi\,,
 \end{eqnarray}
where $\chi$ is a dimensionless measure of the energy-density.
In order to regain general relativity at low energies, the
effective 4D Newton constant is defined by~\cite{T}
\begin{equation}\label{m4}
\kappa_4^2\equiv {8\pi\over M_4^2}= {\kappa_5^4\lambda\over
6(1-4\alpha\Lambda_5/9)}\,.
\end{equation}
When $\alpha=0$, we recover the RS expression. We can fine-tune
the brane tension to achieve zero cosmological constant on the
brane~\cite{T}:
\begin{equation}\label{sig}
\kappa_5^4\lambda^2=-4\Lambda_5+{1\over\alpha}\left[1 -\left(1+
{4\over3} \alpha\Lambda_5\right)^{\!3/2} \right].
\end{equation}
Equations~(\ref{m4}) and (\ref{sig}) may be rewritten
as
 \begin{eqnarray}
\kappa_5^4\lambda &=& 2\kappa_4^2(1+4\alpha\mu^2)
(3-4\alpha\mu^2)\,,\label{m4'} \\ \kappa_5^2\lambda &=&
2\mu(3-4\alpha\mu^2)\,. \label{sig'}
 \end{eqnarray}
These equations imply
\begin{equation}\label{k4k5}
{\kappa_5^2 \over \kappa_4^2}= {1+4\alpha\mu^2 \over \mu}\,.
\end{equation}
The modified Friedman equation~(\ref{mfe}), together with
Eq.~(\ref{chi}), shows that there is a characteristic GB energy
scale\cite{Duf}
 \begin{equation} 
\label{mgb}
M_{\rm GB}= \left[{{2(1-4\alpha\mu^2)^3} \over {\alpha} \kappa_5^4
}\right]^{1/8}\,,
 \end{equation}
such that the GB high energy regime ($\chi\gg1$) is characterized by $\rho+\lambda
\gg M_{\rm GB}^4$. If we consider the GB term in the action as a
correction to RS gravity, then $M_{\rm GB}$ is greater than the RS energy
scale $\lambda^{1/4}$ and this imposes
a restriction on the Gauss-Bonnet coupling $\beta \equiv \alpha\mu^2$\cite{Duf}
 \begin{equation} \label{lim2}
\lambda < M_{\rm GB}^4~\Rightarrow~\beta < 0.038\,.
 \end{equation}
 
Expanding Eq.~(\ref{mfe}) in $\chi$, we find three regimes for the
dynamical history of the brane universe\cite{Duf,ssr,lidsey,T}
:
 \begin{eqnarray}
\rho\gg M_{\rm GB}^4~& \Rightarrow &~ H^2\approx \left[ {\kappa_5^2 \over
16\alpha}\, \rho \right]^{2/3}\,(GB)\,,\label{vhe}\\
M_{\rm GB}^4 \gg \rho\gg\lambda~& \Rightarrow &~ H^2\approx {\kappa_4^2
\over
6\lambda}\, \rho^{2}~~~~~~~~~(RS)\,,\label{he}\\
\rho\ll\lambda~& \Rightarrow &~ H^2\approx {\kappa_4^2 \over 3}\,
\rho~~~~~~~~~~(GR)\,. \label{gr}
\end{eqnarray}
In what follows we shall address the issues of tachyon inflation in the background described by (\ref{mfe}) $\&$
(\ref{chi}). We shall also investigate the specific features of ordinary scalar field dynamics in the extreme regimes given
by (\ref{vhe}), (\ref{he}) $\&$ (\ref{gr}).
\section{Tachyon Inflation on the Gauss-Bonnet Brane}
It was recently suggested that rolling tachyon condensate, in
a class of string theories, might have interesting cosmological consequences.
It was shown by Sen\cite{sen1,sen2} that the decay of  D-branes produces a pressure-less gas
with finite energy density that resembles classical dust. Attempts have been made to construct viable
cosmological model using rolling tachyon field as a suitable candidate for inflaton, dark matter or
dark energy\cite{tachyonindustry}. As for the inflation, the rolling tachyon models are faced with difficulties
related to the requirement of enough inflation and the right level of density perturbations. It seems to be
impossible to meet these requirements if we stick to string theory tachyons.
%as the string inspired effective potentials do not contain any free parameter to ensure enough slow roll and the COBE normalized level of density perturbations. 
In what follows we shall consider the tachyonic potentials in purely phenomenological context
to obtain viable models of inflation.
Unfortunately, even after this relaxation, the tachyonic models face difficulties associated
with reheating\cite{reh1} and the formation of caustics/kinks\cite{reh} and we do not address these problem 
in this paper. We
should, however, note that the model based upon the rolling massive scalar field on $\bar{D}_3$ brane is 
free from these difficulties\cite{gss}, perhaps, except the formation of caustics, which requires further
investigation \footnote{We thank A Starobinsky for his comment on the problem of caustics formation.}.\par

%%%%%%%%%%%%%%%%%%%%%%%%
The tachyonic field is described by the following action
\begin{eqnarray}
S&=& \int
d^4x\biggl\{\sqrt{-g}\left(\frac{R}{2\kappa^2}\right)
\nonumber \\
&-&V(\phi)\sqrt{-\det(g_{ab}+
\partial_{a}\phi\partial_{b}\phi)}\biggr\}\,.
\label{eq0}
\end{eqnarray}

%%%%%%%%%%%%%^M

In a spatially flat
Friedmann-Robertson-Walker (FRW) background, 
the energy momentum tensor which follows from (\ref{eq0}) for the Born-Infeld scalar $\phi$
acquires the diagonal form $T^{\mu}_{\nu}={\rm
diag}\left(-\rho,p,p,p\right)$. The energy density $\rho$
and the pressure $p$, in this case, are given by [we use the signature $(-, +, +,
+)$],
%%%%%%%%%%%%%^M
\begin{eqnarray}
\rho &=& {V(\phi) \over {\sqrt{1-\dot{\phi}^2}}}\,,\\
p &=& -V(\phi)\sqrt{1-\dot{\phi}^2}\,.
\label{rhop}
\end{eqnarray}
The equation of motion of the rolling scalar field follows
from Eq.~(\ref{eq0}) 
%%%%%%%%%%%%%^M
\begin{eqnarray}
{\ddot{\phi} \over {1-\dot{\phi}^2}}+3H \dot{\phi}+{V_{\phi}
\over V({\phi})}=0\,, \label{ddphi0}
\end{eqnarray}
which is equivalent to the conservation equation
\begin{equation}
{\dot{\rho} \over \rho}+3H(1+w)=0
\label{conseq}
\end{equation}
We now describe inflation on the brane assuming slow roll approximation, $\dot{\phi}^{2} << V$ and $|\ddot{\phi}| << H |\dot{\phi}|$. The energy density becomes $\rho \sim V(\phi)$ and using Eqs. (\ref{chi}) and
(\ref{sig}) we obtain (for the weak GB coupling defined by (\ref{lim2}))
\begin{equation}
V \simeq \sqrt{ \left( \frac{\lambda  }{3 \alpha \kappa_{4}^{2}} \right)} \; \sinh \chi .
\label{hpot1}
\end{equation}
The slow-roll parameters in this case become
\begin{eqnarray}
\epsilon& =& \nonumber  \left( \frac{2 \lambda}{\kappa_{4}^{2}} \frac{V_{\phi}^2}{V^4} \right) \; \epsilon_{GB} \,, \\
\eta &=& \left( \frac{2 \lambda} {\kappa_{4}^{2}V^2}  \left(  \ln V \right)_{\phi \phi} \right) \; \eta_{GB}\,,
\label{slow1}
\end{eqnarray}
where the GB corrections to the RS values are given by
\begin{eqnarray}
\epsilon_{GB}& =& \nonumber
\left[ \frac{2 }{27} \frac{\sinh (2\chi/3) \tanh \chi \sinh^2 \chi}{(  \cosh(2\chi/3) - 1 )^2} \right]\,, \\
\eta_{GB}& =&\left[ \frac{2 }{9 } {\sinh^2\chi \over { \cosh(2\chi/3) - 1)}} \right]\,.
\end{eqnarray}
The number of e-folds of inflationary expansion, ${\cal N} = \int H dt $, is obtained using (\ref{mfe}) and (\ref{ddphi0}), which is given by
\begin{equation}
{\cal N} = 3\int_{\chi_e}^{\chi_N}H^2 \frac{V}{V_{\chi}}
\left(\frac{{\rm d}\phi}{{\rm d}\chi}\right)^2d\chi\,,
\end{equation}
which using Eqs. (\ref{ddphi0}) and (\ref{hpot1}) takes the form
\begin{equation}
\label{efold} 
{\cal N}(\chi) = - \frac{3}{4 \alpha} \int_{\chi_N}^{\chi_{end}} d\chi \left( \frac{d \phi}{d \chi} \right)^2 \left(  \cosh (2\chi/3) - 1 \right) \tanh\chi 
\end{equation}
We should note that we have used the weak coupling nature of GB correction while writing Eqs. (\ref{hpot1}), (\ref{slow1})
and (\ref{efold}).
\subsection{Inflation with  Polynomial Type Potential }
We shall now assume that the potential for Born-Infeld scalar field is
%%%%%%%%%%%%%%%%%^M
\begin{equation}
\label{po}
V(\phi)=V_0\phi^p\,,
\end{equation}
%%%%%%%%%%%%%%%%%^M
where $V_0$ and $p$ are constants.
We are mainly interested in the cases of 
$p=2$ (massive inflaton), $p=4$ (massless inflaton)
and $p \to \infty$ (exponential potential). 
For the potential (\ref{po}) two slow-roll parameters 
can be written as
\begin{eqnarray}
\label{epsilon2}
\epsilon &=&\nonumber 
\frac{4\lambda p^2V_0^{2/p}}
{27\kappa_4^2A^{2(p+1)/p}}
f(\chi)\,, \\
\eta &=&-
\frac{4\lambda pV_0^{2/p}}
{9\kappa_4^2A^{2(p+1)/p}}g(\chi) \,,
\label{eta2}  
\end{eqnarray}
where $A=\sqrt{3\lambda/\alpha \kappa_4^2}$ and $f(\chi)$, $g(\chi)$ are given by
\begin{eqnarray}
f(\chi)&=& \nonumber \frac{\sinh (2\chi/3) \, \tanh \chi \, 
(\sinh \chi)^{-2/p}}{\left[
\cosh (2\chi/3) -1 \right]^2} \,, \\
g(\chi)&=&
\frac{(\sinh\chi)^{-2/p}}{(\cosh (2\chi/3) -1)} \,.
\end{eqnarray}
A comment on the behavior of slow roll parameters is in order. As pointed out in Ref.\cite{ssr}, both $\epsilon$ and $\eta$
exhibit a peculiarity for $p=6$ (in case of ordinary scalar field inflation) in the region $\chi<<1$: they are increasing functions of $\chi$ for $p<6$ whereas the
situation is reversed for $p>6$. It turns out that $p=6$ case, also gets distinguished for large values of $\chi$, i.e, 
in the
GB regime where the dynamics is described by a simple equation $H^2 \sim \rho^{2/3}$. Indeed, in region  $\chi>>1$, the
the slow roll parameters behave as
\begin{eqnarray}
\label{epasim}
\epsilon,\eta && \propto \chi^{(p-6)/3p}~~~~~~~(Ordinary~ scalar~ field)\,,\\
\epsilon,\eta  && \propto \chi^{-(3+p)/3p}~~~~~~(Tachyon~ field) \,.
\label{etaasim}
\end{eqnarray}
It is clear from Eqs. (\ref{epasim}) $\&$ (\ref{etaasim}) that in the GB regime, the slow roll
parameters exhibit a specific behavior in case $p=6$ for ordinary scalar field whereas the similar behavior
is realized for tachyon field if $p=-3$\cite{calcagni} (see Ref.\cite{others} which deals with similar problem
in case of RS and standard GR). 
%%%%%%%%%%%%%%%%%%%%%%%%%%%%%%%%%%%%%%%%%%%%%%%%%%%%%%%%%%%%%%%%%%%%%%%
After a brief remark on the scalar field dynamic in patch cosmology, we return to the 
full dynamics described by (\ref{mfe}) $\&$ (\ref{chi}). We now compute the number of inflationary e-foldings for polynomial potential (\ref{po}) 
\begin{equation}
{\cal N} = - \frac{3 V_0^{-2/p}}{8 \alpha p^2 A^{-2/p}} \int_{\chi_{N}}^{\chi_{end}} {d\chi  \frac{\left[  \cosh (2 \chi /3) - 1 \right]}{\left(\sinh(\chi)\right)^{2(p-1)/p}} \sinh(2 \chi)}.
\label{N2}
\end{equation}
For a general $p$, it is not possible to get a close analytical expression for ${\cal N}$. However, for particular values
$p \pm 2 ,\infty$, the integral in (\ref{N2}) can be computed analytically. For one of the values of interest $p=4$, we shall opt
for the numerical computation of the integral. It will be instructive to present the expression for the number of e-foldings
, in general, as follows
\begin{equation}
{\cal N}=\frac{3}{8\alpha p^2V_0^{2/p}A^{-2/p}}
\left[ F(\chi) \right]_{\chi_e}^{\chi_N}
\label{N3}
\end{equation}
In order to estimate the maximum number of e-foldings, we can assume that inflation ends in the RS regime ($\chi<<1$) which
allows us to write (\ref{N3}) as
\begin{equation}
{\cal N} =\frac{3}{8\alpha p^2V_0^{2/p}A^{-2/p}}
\left[F(\chi_N)-\frac{2p}{9(1+p)}
\chi_e^{2(p+1)/p}\right]
\label{N4}   `
\end{equation}
We observe that the slow roll parameter $\epsilon$ scales as $ \chi^{2(p+1)/p}$ for $\chi <<1$ which helps to 
estimate the value $\chi_e$ at the end of inflation
\begin{equation}
\chi_e^{2(p+1)/p}=\frac{2\lambda p^2 V_0^{2/p}}{\kappa_4^2 A^{2(p+1)/p}}
\label{chi1}
\end{equation}
Using Eqs. (\ref{N3}) and (\ref{chi1}), we can express $\chi_e$ through $F(\chi_N)$ as
\begin{equation}
\chi_e^{2(p+1)/p}=\frac{9(p+1)}{2\left[2{\cal N}(p+1)+1\right]} F(\chi_N) \,,
\label{chif}
\end{equation}
which for $p \rightarrow \infty$ reduces the expression for $\chi_e$ obtained in Ref\cite{paul}. for exponential potential. We now give the analytical expressions for the function
$F$ for $p=\pm 2,\infty$
\begin{eqnarray}
F(\chi)&=&\nonumber {4\over 5}{(6\cosh(2\chi/3)-1)}\sinh^3(\chi/3)~~~~~(p=2)\,, \\
F(\chi)&=&\nonumber 3\cosh(2\chi/3)-\ln(1+2\cosh(2\chi/3)\\
&+& \nonumber 2\ln (\sinh(\chi/3)-2\ln(\sinh(\chi))\\
&+&  \nonumber 3(\ln(3)-1)~~~~~~~~~~~~~~~~~~~~~~~~(p=\infty)\,, \\
F(\chi)&=&\nonumber 4{ {ArcTan\left(2\sinh(\chi/3)/\sqrt{3}\right)} \over {\sqrt{3}}} \\
&-& \frac{4 \sinh(\chi/3)}{1+2 \cosh(2\chi/3)}~~~~~~~~~~~~(p=-2) \,,
\end{eqnarray}
whereas for other values of $p$, the function $F(\chi)$ should be
evaluated numerically.

\begin{figure}
\resizebox{3.5in}{!}{\includegraphics{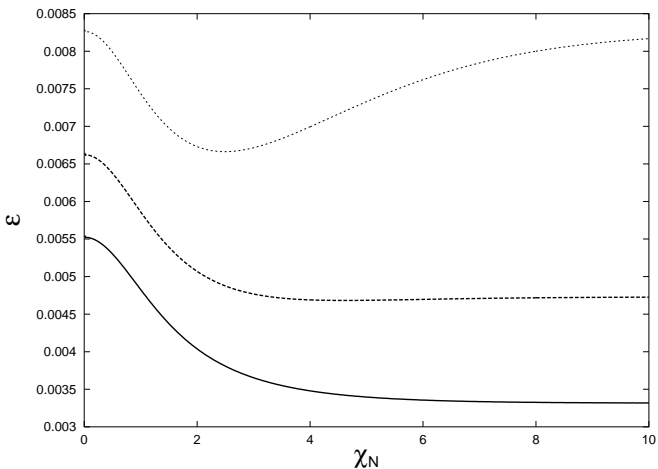}}
%\small{
\caption{The slow roll parameter $\epsilon$ as a function of $\chi_N$ for the number of
e-folds ${\cal N}=60$ in case of the potential (\ref{po}). The solid line corresponds to $p=2$, 
dashed and the dotted lines corresponding to $p=4$ and the exponential potential respectively.}
\label{epsilon}
\end{figure}

\begin{figure}
\resizebox{3.5in}{!}{\includegraphics{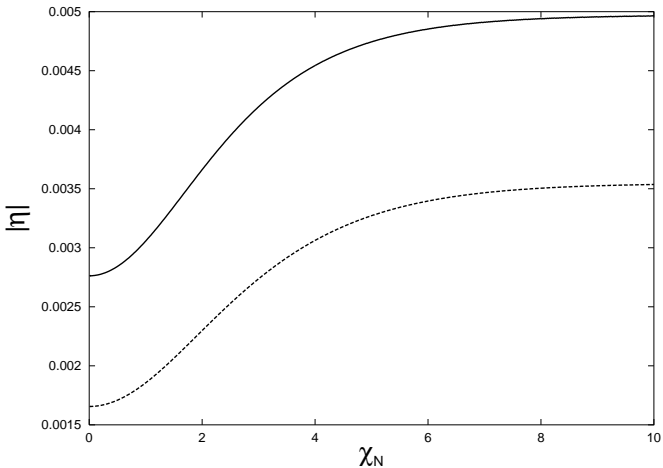}}
%\small{
\caption{Plot of $\eta$ as a function of  $\chi_N$ for the number of
e-folds ${\cal N}=60$. The dashed line corresponds to $p=4$ whereas the solid line to $p=2$;
$\eta \equiv 0$ in case of the exponential potential.}
\label{eta}
\end{figure}

The slow roll parameters can now be cast entirely as a known function of $\chi_N$
\begin{eqnarray}
\label{epsilonF}
\epsilon&=&\nonumber \frac{(p+1)F(\chi_N)}{3\left[2{\cal N}(p+1)+p\right]}\left(\frac{\sinh (2\chi_N/3) \tanh \chi_N 
(\sinh \chi_N)^{-2/p}}{\left[
\cosh(2\chi_N/3) -1 \right]^2}\right)\,, \\
\eta&=&-\frac{(p+1)F(\chi_N)}{p\left[2{\cal N}(p+1)+p\right]}\left(\frac{(\sinh\chi_N)^{-2/p}}{(\cosh (2\chi_N/3) -1)}\right)\,
\end{eqnarray}
which for $p \rightarrow \infty$ corresponds to the case of exponential potential; the slow roll parameter $\eta$ vanishes
in this limit and (\ref{epsilonF}) reduces to the expression obtained in Ref\cite{paul}.\\
As mentioned above, the cases corresponding to $p \pm 2$ and $p=\infty$ (exponential potential)
can be treated analytically. In case of $p=4$, we get complicated combinations of hyper-geometric functions; it is not very illuminating to produce theme in the text and we we have studied this case numerically. We have ensured that
the numerics in case of  $p=\pm 2$ and $p=\infty$ produces our analytical results. In Figs. \ref{epsilon}
$\&$ \ref{eta}, we have plotted the slow roll parameters $\epsilon$ and $\eta$ for three cases. We
observe that for large values of $p$, the slow roll parameter $\epsilon$ has minimum in the
intermediate region which increases and approaches a constant value as we move towards the GB regime (large values of $\chi_N$). The minimum becomes less and less pronounced for smaller values of $p$.
The slow roll parameter $\eta \equiv 0$ in case of the exponential potential, whereas in other two cases,
it represents a monotonically increasing function of $\chi_N$ approaching a constant value in the GB regime (see Fig. \ref{eta}). It is interesting to compare these features with GB inflation in case of ordinary
scalar field. In the later case, the slow roll parameters are monotonously increasing function of
$\chi_N$ for large values of $p$ in contrast to the GB tachyonic inflation where they assume a minimum value in the intermediate region
and then gradually approach a constant value. 
Secondly, numerical values of these parameters, at all energy scales and for $\forall p >0$, remain much smaller than their counter parts associated with ordinary scalar field GB inflation.

%%%%%%%%%%%%%%%%%%%%%%%%%%%%%%%%%%%%%%%%%%%%%%%%%%%%%%%%%%%%%%%%%%%%%%%%%%%%%

\subsection{Perturbation from Gauss-Bonnet Inflation}
\begin{figure}
\resizebox{3.5in}{!}{\includegraphics{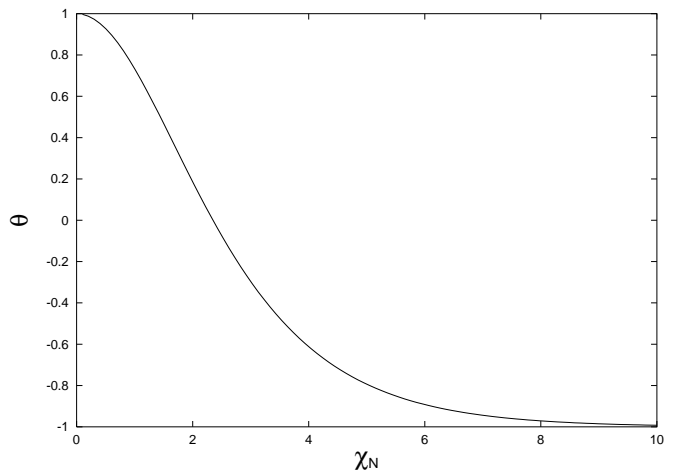}}
%\small{
\caption{Plot of function $\theta$ versus the energy scale $\chi_N$. The function interpolates between $1$ and $-1$ as $\chi_N$
runs from RS region to GB regime.}
\label{figtheta}
\end{figure}

Hwang and Noh \cite{Hwang} provided
the formalism to evaluate the perturbation spectra for the
general action
%%%%%%%%%%%%%
\begin{eqnarray} 
S=\int {\rm d}^4 x\sqrt{-g}\frac12 f(R, \phi, X)\,, 
\end{eqnarray}
%%%%%%%%%%%%%
which includes our action (\ref{eq0}). Here the function $f$
depends upon the Ricci scalar $R$, a scalar field $\phi$ and its
derivative $X=(\nabla \phi)^2/2$.
% [we use the signature $(-, +, +,
%+)$].
The Born-Infeld scalar field corresponds to the case with
%%%%%%%%%%%%%
\beqa f=\frac{R}{\kappa^2}+2\Lambda-2V(\phi)\sqrt{1+2X}\,.
\label{tach} \eeqa
%%%%%%%%%%%%%
The amplitude of density perturbations in this case is given by\cite{Hwang}
%%%%%%%%%%%%%
\begin{eqnarray}
A_{\rm S}^2=\left(\frac{H^2}{2\pi^2\dot{\phi}} \right)^2
\frac{1}{Z_{\rm S}}\,, \label{powersca}
\end{eqnarray}
where $Z_{\rm S}\equiv -(f_{,X}/2+f_{,XX}X)=
V(1-\dot{\phi}^2)^{-3/2}$. Under the slow-roll approximation,
 the power spectrum of curvature
perturbations is estimated to be \cite{Hwang}
\begin{equation}
A_{\rm S}^2=\left(\frac{H^2}{2\pi\dot{\phi}} \right)^2 {1 \over V}
\label{As}
\end{equation}
The extra piece of $V$ occurring in (\ref{As}) leads to the modified expression for spectral index $n_S$ in
case of tachyon field
\begin{equation}
n_S-1=\frac {d \ln A_S^2}{d\ln k}|_{k=aH}=-(4+\theta(\chi))\epsilon+2\eta,
\label{ns}
\end{equation}
%%%%%%%%%%%%%
where $\theta(\chi)$ is given by
\begin{eqnarray}
\theta(\chi)&=& \nonumber 2\left(1-{{3\cal{G}(\chi)} \over 2}\right) \,, \\
\cal{G}(\chi)&=&\frac{\left(\cosh(2\chi/3)-1 \right)}{\sinh(2\chi/3)}\coth(\chi) \,
\label{theta}
\end{eqnarray}
We have used Eqs. (\ref{mfe}) $\&$ (\ref{hpot1}) in deriving (\ref{theta}). The function $\theta(\chi)$ encodes
the GB effects for tachyon inflation. It interpolates between $1$ and $-1$ (${\cal G}(\chi)$ 
varies from zero to $1$) as  $\chi$ varies from
RS to GB limit (Fig. \ref{figtheta}) which is in confirmation with the findings of Refs.\cite{calcagni,gs} in extreme limits.

The tensor perturbations in brane world with Gauss-Bonnet term in the bulk were recently studied in Ref.\cite{Duf}. The 
 amplitude of tensor perturbations was shown to be given by
%%%%%%%%%%%%%

%%%%%%%%%%%%%

\begin{eqnarray}
A_T^2=\left[\kappa^4{{H^2} \over {4\pi^2}}\right]{\cal F}_\beta^2(H/\mu)\,,
\end{eqnarray}
where the function ${\cal F}_{\beta}$ contains the information about the GB term
\begin{equation}
{\cal F}_\beta^{-2}=\sqrt{1+x^2}-\left({{1-\beta}\over {1+\beta}}\right)\sinh^{-1}x^{-1}~~~~~~~(x \equiv H/\mu).
\end{equation}
The dimensionless variables $x$ and $\chi$ associated with energy scales are related to each other via
the  Eqs. (\ref{mfe}) $\&$ (\ref{chi}).\\
The tensor spectral index in this  case is
\begin{eqnarray}
n_T={{d\ln A_T^2}\over {d\ln k}}|_{k=aH}=-\epsilon G_\beta(x) \,,
\label{nt}
\end{eqnarray}
where $G_{\beta}(x)$ is given by
\begin{equation}
G_\beta(x)=1- \frac{x {\cal F}_\beta^2\left(1-(1-\beta)\sqrt{1+x^2}\sinh^{-1}x^{-1}\right)}
           {(1+\beta^2)\sqrt{1+x^2)}}
\label{G}          
\end{equation}
The tensor to scalar ratio is defined as
\begin{equation}
R=16{A_T^2 \over A_S^2}
\end{equation}
Following Ref\cite{Duf}, we have the expression for the tensor to scalar ratio $R$
\begin{eqnarray}
R&=&-\nonumber 8 Q(x) n_T\,, \\
Q(x)&=&\left({{1+\beta+2\beta x^2}\over {1+\beta+\beta x^2}}\right)\,,
\label{Q}
\end{eqnarray}
where $Q$ carries the information of GB correction. It determines the size of breaking of
degeneracy of the consistency relation in Gauss-Bonnet brane world inflation.
We finally express the ratio of perturbations through the spectral index
using Eqs. (\ref{ns}), (\ref{nt}) and (\ref{Q}) as
\begin{equation}
R={\cal D}(\chi_N)\left[\frac {8}{4+\theta}(1-n_S)+\frac{16}{(4+\theta)} \eta\right] \,,
\end{equation}
where ${\cal D}(\chi_N)=Q(\chi_N)G_\beta(\chi_N)$.
Knowing the slow roll parameters and the functions ${\cal D}(\chi_N)$ $\&$ $\theta(\chi)$, we can evaluate the spectral index
$n_S$ and the ratio $R$. In Figs. \ref{figns} $\&$ \ref{figR}, we have displayed their 
dependence on the dimensionless energy scale $\chi_N$. The spectral index rises to 
maximum in the intermediate region and then gradually decreases approaching a constant value in GB regime ($\chi_N>>1$). It
improves in general for lower values of $p$ ($p >0 $).
In case of the exponential potential, the maximum value of the spectral index is nearly equal to $0.97$ for ${\cal N}=60$
which is consistent with the result obtained earlier in \cite{paul}. \\
%%%%%%%%%%%%%%%%%%%%%%%%%%%%%%%%%%%%%%%%%%%%%
\subsection{Asymptotic Scale Invariance in GB Tachyon Inflation}
\begin{figure}
\resizebox{3.5in}{!}{\includegraphics{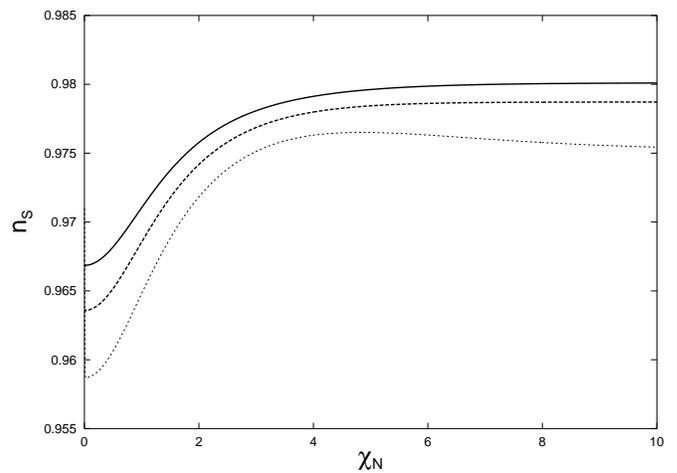}}
%\small{
\caption{Plot of the spectral index $n_S$ versus the dimensionless energy scale $\chi_N$ for
the number of e-folds ${\cal N}=60$. Solid line corresponds to $p=2$, dashed to $p=4$ and
dotted line to $p=\infty$ (exponential potential).}
\label{figns}
\end{figure}

\begin{figure}
\resizebox{3.5in}{!}{\includegraphics{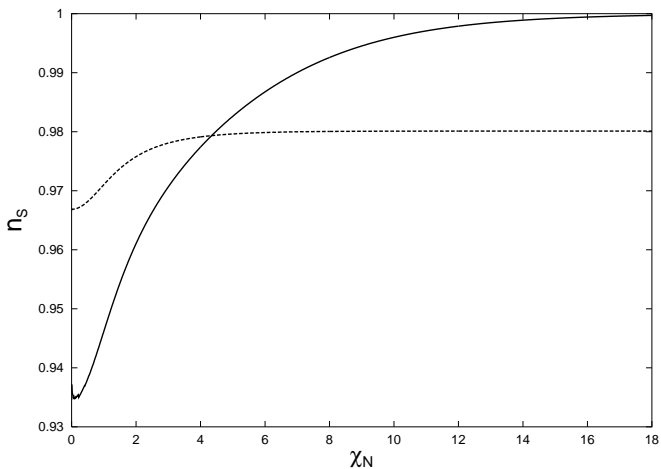}}
%\small{
\caption{Spectral index $n_S$ versus the dimensionless energy scale $\chi_N$ for
the number of e-folds ${\cal N}=60$ in case of potential $V \sim \phi^p$. Solid line corresponds to $p=-2$, 
dashed to $p=2$.
The spectral index for inverse square potential is seen approaching the scale invariance limit ($n_S=1$) in the GB regime.}
\label{fignspboth}
\end{figure}

\begin{figure}
\resizebox{3.5in}{!}{\includegraphics{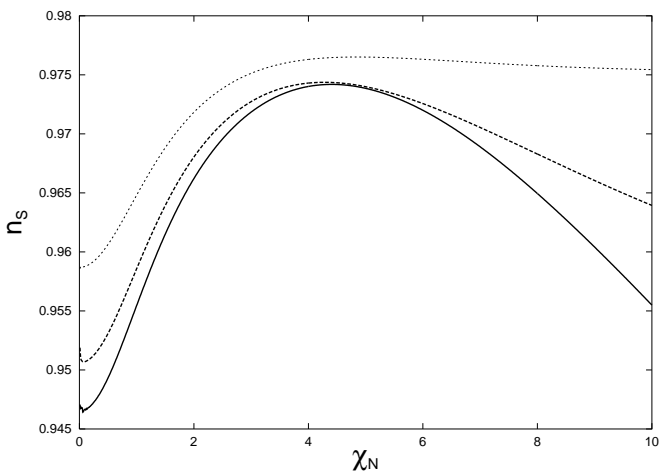}}
\caption{Plot of spectral index $n_S$ versus  energy scale $\chi_N$ for
the number of e-folds ${\cal N}=60$ in case of potential $V \sim \phi^p$ with $p=-3$ (solid line) and $p=-4$ 
(dashed) line. The dotted line corresponds to exponential potential.}

\label{fignsrunpot}
\end{figure}

As seen in Fig. \ref{figns} , the spectral index $n_S$ improves for lower values of the exponent $p$.
It would really be interesting to compare this situation with the standard inflationary
scenario in presence of the GB correction in the bulk. In this case,
the spectral index shows a very different behavior relative to the tachyonic GB inflation for exponential potential.
It monotonously increases and approaches $1$ for large $\chi_N$\cite{lidsey}. Actually, the exponential potential is
special to ordinary GB inflation which is related to the fact that scale invariance is exact in this case if the background dynamics is governed by
the Hubble equation $H^2 \sim \rho^{2/3}$\cite{lidsey}. And this is certainly not true for tachyon field as it is governed by different
dynamics. Interestingly, exact scaling for tachyon GB inflation is realized by a field potential very different
from the exponential function. Indeed, let us consider the slow roll parameters in
the background
described by $H^2 \sim \rho^{q}$
\begin{eqnarray}
\epsilon&=& \nonumber \frac{q}{6H^2}\left({V_{,\phi} \over V} \right)^2 \,, \\
\eta&=& \frac{1}{3 H^2} \left(\ln V \right)_{,\phi \phi} \,,
\label{slowq}
\end{eqnarray}
which for the power law type of potential $V \sim \phi^p$ leads to the following expression for the spectral index $n_S$ in the asymptotic limit $\chi_N>>1$
\begin{equation}
n_S-1=- \frac{1}{3 H^2}\left(\frac{(4+\theta) pq}{2}+2 \right){p \over \phi^2}
\label{scaling}
\end{equation}
In deriving Eq. (\ref{scaling}), we have used Eq. (\ref{ns}). 
It should be noted that the general expressions of slow roll parameters (\ref{slow1}) reduce to 
(\ref{slowq}) in the  limits of small $\chi$ with $q=2$ and large $\chi$ with $q=2/3$ and that Eq. (\ref{scaling})
is valid in the asymptotic regimes  $\chi_N<<1$ (RS $-$ regime) and $\chi_N>>1$ (GB $-$ regime).
%%%%%%%%%%%%%%
%%%%%%%%%%%%%%%%%%%%%%%%%%%%%%%%
For scale invariance of spectrum, the (RHS) of (\ref{scaling}) should vanish leading to the simple relation
\begin{eqnarray}
p=-\frac{4}{q(4+\theta)} \,,
\end{eqnarray}
which gives rise to $p=-2$~for GB patch ($q=2/3$) and $p=-2/5$ in case of RS patch ($q=2$), in agreement with the result obtained in\cite{gs}. Here we have taken into account that $\theta(\chi) \to \pm 1$ in the limits of
$\chi_N<<1$ and $\chi_N>>1$ respectively.
Our treatment
of the full dynamics confirms this feature in the high energy GB regime (see, Fig.\ref{fignspboth}).\\
We have also considered models corresponding to larger inverse powers than the inverse square potential.
We find that the numerical values of $n_S$ for $p\le -3$ are lower as compared to the case of an exponential potential (Fig. \ref{fignsrunpot}) and approach the later in the limit of large negative $p$. The
crossing takes place for $p > -3$ allowing the scale invariant limit to be reached for $p=-2$.

%%%%%%%%%%%%%%%%%%%%%%%%%%%%%%%%%%%%%%%%%%%
\subsection{Tensor to Scalar Ratio of Perturbations $R$}
\begin{figure}
\resizebox{3.5in}{!}{\includegraphics{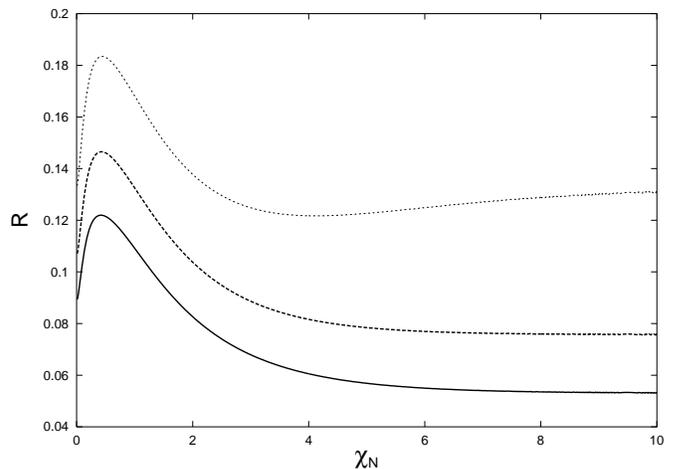}}
%\small{
\caption{The tensor to scalar ratio of perturbations $R$ is shown as a function
of the dimensionless scale $\chi_N$ for ${\cal N}=60$. The solid line corresponds
to the case of $p=2$. The dashed and dotted lines corresponding to $p=4$ and the exponential potential.}
\label{figR}
\end{figure}

\begin{figure}
\resizebox{3.5in}{!}{\includegraphics{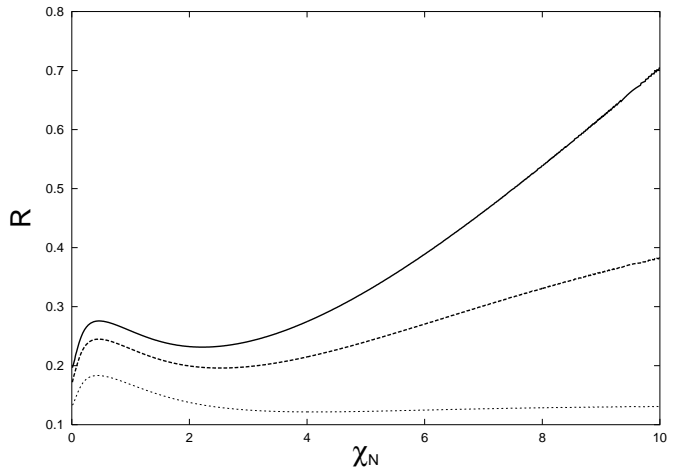}}
%\small{
\caption{The energy scale dependence of $R$ 
for the model described in Fig. \ref{fignsrunpot}. The solid line corresponds
to the case of $p=-3$, dashed and dotted lines corresponding to $p=-4$ and  exponential potential.
$R$ takes minimum value in the intermediate region between RS and GB regimes.}
\label{figRrunpot}
\end{figure}

\begin{figure}
\resizebox{3.5in}{!}{\includegraphics{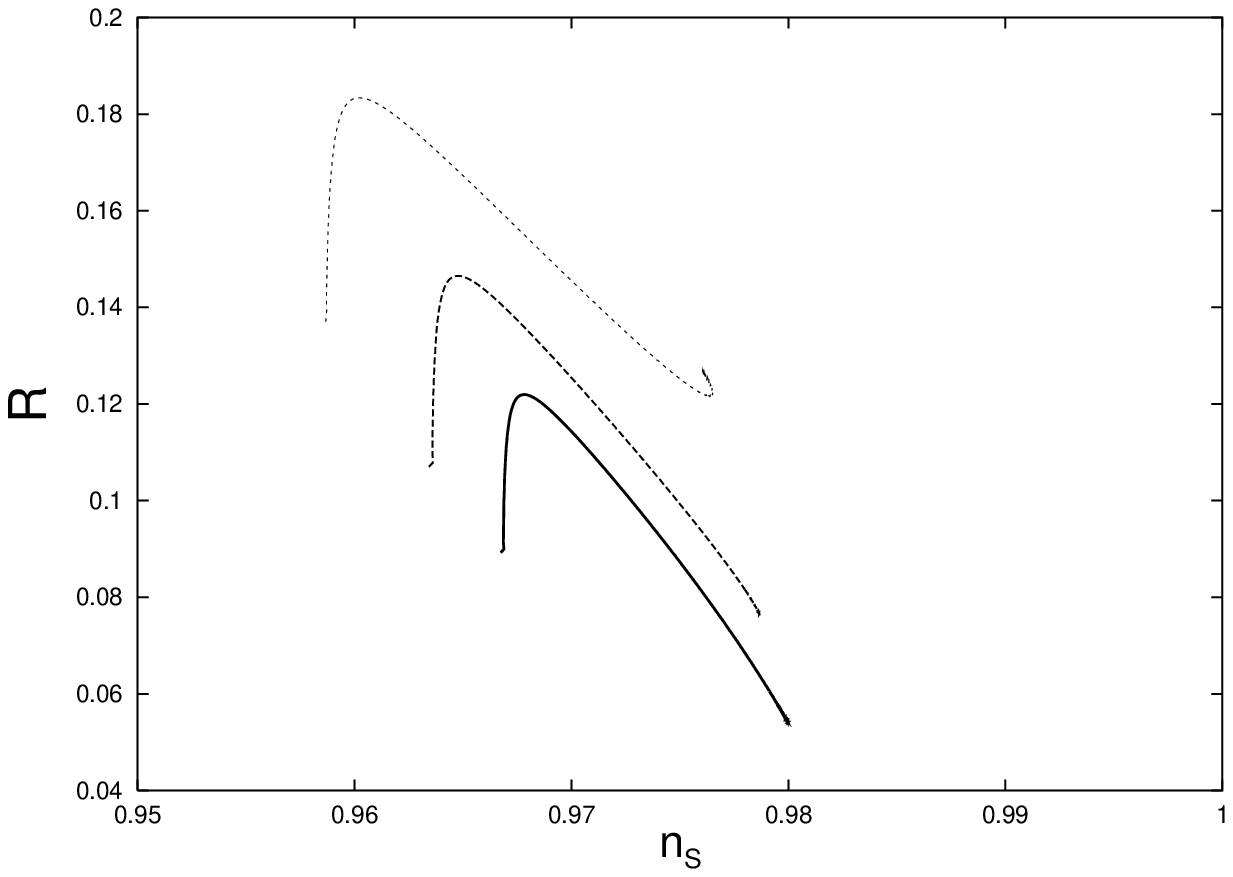}}
%\small{
\caption{The tensor to scalar ratio of perturbations $R$ is shown on the ($R,n_S$) plane for ${\cal N}=60$. The solid line corresponds to the case of $p=2$. The dashed and dotted lines corresponding to $p=4$ and the exponential potential.}
\label{figRns}
\end{figure}

The behavior of the tensor to scalar ratio of perturbations is dictated by the features possessed by the functions ${\cal D}(\chi_N)$
and $n_S$. The ratio $R$ is plotted in Fig. \ref{figR}. The function $R$ peaks around the RS regime which
subsequently decreases to minimum and increases thereafter approaching a constant value in the GB regime. This is a very
important feature of GB inflation common to both tachyonic as well as non-tachyonic models. The RS value of the ratio $R$ is 
generally larger relative to the case of GR\cite{gs}. The minimum
of the function $R$ is attributed to the fact that while passing from RS regime characterized by $H^2 \sim \rho^2$
to the high energy GB limit with $H^2 \sim \rho^{2/3}$, there is an intermediate region which mimics the GR like
behavior. In case of lower values of $p$, the minimum of $R$ is not distinguished. 
The numerical values of $R$ as a function of $\chi_N$ are generally smaller for less steeper potentials. 
We find that the tensor to scalar ratio of
perturbations is very low for all the values of the exponent $p>0$ at all the energy scales 
thereby providing support to the recent analysis of Ref.\cite{gs}
in the limiting cases. The tachyonic model of inflation with polynomial type of potentials
is within the $1\sigma$ contour bonds at all energy scales for $p>0$ (see Fig. \ref{figRns} and the observational contours
given in Ref.\cite{ssr}; also see Ref\cite{pi} on the related theme). In case of the run away potentials for small negative values of $p$, the tensor
to scalar ratio becomes large for large values of $\chi_N$ and it is suppressed in the intermediate region (Fig. \ref{figRrunpot}). Thus, there is a possibility for these
models to be consistent with observation in the intermediate region between RS and GB which is
analogous to ordinary scalar field GB inflation with steep potentials\cite{ssr}.
Finally, we should remark that the dimensionless density scale can not increase indefinitely, it is restricted
by the quantum gravity limit which corresponds to $\rho <\kappa_5^{-8/3}$
\begin{equation}
\frac{\alpha^3\lambda}{\kappa_4^2}>48\sinh^6(\chi_N)
\label{cons}
\end{equation}
Using Eqs. (\ref{As}), (\ref{chi1})  $\&$ (\ref{chif}) along with
COBE normalized value of density perturbations, we can express  
$\alpha^3\lambda/\kappa_4^2$, entirely, as a function of energy scale $\chi_N$ and the number of
e-folds $\cal{N}$. The constraint (\ref{cons}), then leads to
an upper bound on the variable $\chi_N$. In case of ordinary scalar field GB inflation, it was very important
to find these bounds as the tensor to scalar ratio $R$, in general, is a monotonously increasing function
of $\chi_N$ which becomes large in high energy GB regime. In our case, as mentioned above, the ratio
remains very low for all values of $\chi_N$ in case of any generic positive value of $p$. However, it is true that it makes sense to consider only those values
of the energy scale which are consistent with (\ref{cons}). The upper bounds on $\chi_N$  in our model lies
between $6$ $\&$ $7$ in these cases. 
\begin{figure}
\resizebox{3.5in}{!}{\includegraphics{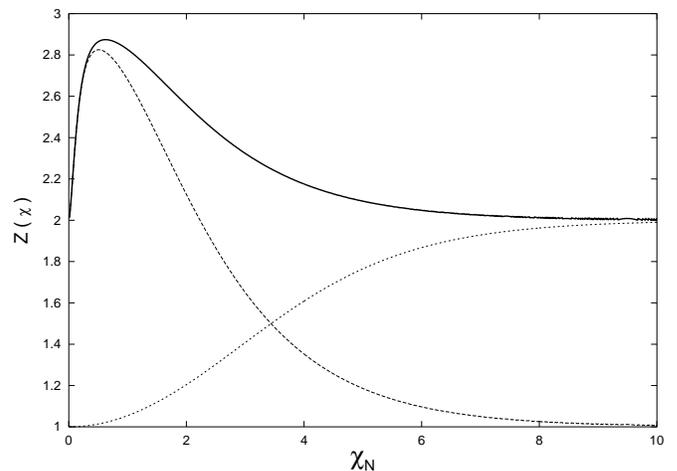}}
%\small{
\caption{Evolution of function $Z= Q $ (dotted line), G (dashed line), ${\cal D}$ (solid line)
with the energy scale $\chi_N$. The degeneracy factor $Q$ evolves from $1$ to $2$ as $\chi_N$ varies from RS ($\chi_N <<1$)
to GB regime, $\chi_N>>1$ ($Q=1$ in the standard GR case corresponding to $\beta=0$). $G$ interpolates
between $2$ and $1$, it expresses the variation of the ratio $n_T/\epsilon$ as the energy scale changes from lower to higher values.The function ${\cal D}$ peaks around the RS regime and
tends to a constant value for large $\chi_N$.}
\label{figQGD}
\end{figure}

%%%%%%%%%%%%%%%%%%%%%%%%%%%%%%%%%%%%%%%%%%%
%%%%%%%%%%%%%%%%%%%%%%%%%%%%%%%%%
%%%%%%%%%%%%%%%%%%%%%%%%%%%%%%
%\small{
%%%%%%%%%%%%%%%%%%%%%%%%%%%%%%%%%%%%%%%%
%%%%%%%%%%%%%%%%%

%%%%%%%%%%%%%%%%%%%%%%%%n

\section{Issues of Scalar Field Dynamics in $H^2 \sim \rho^q$ Cosmology}
So far, three particular models of the form $H^2 \sim \rho^q$
have been considered in the literature. These include: standard
cosmology($q=1$), the Randall-Sundrum brane ($q=2$) and the Gauss-Bonnet
brane ($q=2/3$). The scalar field dynamics in these three cases exhibits 
several important differences. In order to understand the connections between
the power index $q$ in the generalized Friedmann equation and
particular properties of corresponding scalar field dynamics it is
necessary to examine the problem in the general
cosmological background. The general description of the dynamics seems to be
possible
in a number of interesting physical situations.
In what follows  
we shall describe asymptotic behavior of ordinary scalar field near a cosmological
singularity and investigate the  possibilities for the 
existence of scaling solutions for the standard as well as the tachyon field. 

%%%%%%%%%%%%%%%%%%%%%%%%%%%%%%%%%%%%%%%%%%%%%%%%%%%%%%%%%%%%%%%%%%%%i 
\subsection{Asymptotic behavior near  singularity} 
An interesting example of different dynamics in the standard and  
brane cosmologies is related to the behavior of the scalar field near a  
cosmological singularity. It is known that in the standard case the 
scalar  field diverges near a singularity \cite{Foster} while it 
remains finite  in the brane world \cite{ttu}. We shall consider the 
behavior of scalar field near singularity and study the asymptotic
solutions
in a cosmological background governed by $H^2 \sim \rho^q$. Considering 
this problem in the general case  we start with a massless field. 
The equation of motion  
  \begin{equation}
  \ddot\phi + 3 H \dot\phi =0,
 \end{equation}  
  in the background described by $H^2 \sim \rho^q$ gives  
  $ H \sim \dot\phi^q $ which leads to  
  \begin{equation}
 \ddot\phi + \dot\phi^{1+q} =0.
 \label{toporphieq}
 \end{equation}  
 Eq. (\ref{toporphieq}) easily integrates and gives  
  \begin{equation}
 \phi = A (t-t_0)^{1-1/q},
 \label{asympt}
 \end{equation}
 where $A$ and $t_0$ are constant of integration.  
  We observe that the standard cosmology, $q=1$ 
(in this case we can not use (\ref{asympt}) for which the asymptotic 
has the known form $\phi \sim ln(t/t_0)$), is an exceptional 
case which divides all possible asymptotic in two classes. 
For $q < 1$ both $\phi$ and $\dot\phi$ diverge near a cosmological 
singularity. The GB brane belongs to this class with the asymptotic 
$\phi \to 1/\sqrt{t-t_0}$. On the contrary, $q > 1$ leads to 
nonsingular $\phi$ and singular $\dot\phi$ ($\dot\phi$ can not be 
nonsingular because the power index in (\ref{asympt}) is always 
less then unity). The well known example of this dynamics is provided by the
Randall-Sundrum brane with $\phi \to \sqrt{t-t_0}$.\\
It is known that in the standard case the scalar field potential
$V(\phi)$ is not important during the cosmological collapse unless
  it is steeper than exponent (see \cite{Foster2} for detail). In the 
 general case the role of potential  depends on the sign of $q-1$.
  For $q>1$ the asymptotic $\phi \to const$, $\dot\phi \to \infty$
  prevents the potential from plying an important role in the cosmological
  collapse. If $q<1$, however, steep enough potential would destroy
  the regime (\ref{asympt}). For the scalar field growing as
 \begin{eqnarray} 
  \phi \sim (t-t_0)^{-a} \,,
  \label{eq1}
  \end{eqnarray}
  the kinetic energy behaves as
  \begin{equation}
  \dot\phi^2 \sim(t-t_0)^{-2-2a}.
  \label{eq2}
 \end{equation}
  Assuming the power law form of potential $V(\phi) \sim \phi^b$ and using Eqs. (\ref{eq1}) $\&$ (\ref{eq2}), it is easy to see that the potential becomes
  important provided that
\begin{equation}
  b>\frac{2(1+a)}{a}.
 \end{equation}
 Using Eq. (\ref{asympt}) we then find the critical value of the power
  index in the potential 
 \begin{eqnarray} 
  b=\frac{2}{1-q} \,.
  \label{powerlaw}
 \end{eqnarray}
 
 For steeper potentials it is impossible
  to neglect $V(\phi)$ which makes  the asymptotic (\ref{asympt}) invalid and the scalar field
  in a contracting Universe
  enters into a regime of oscillations, similar to described in \cite{Foster2}.\\
  We should emphasize that (\ref{powerlaw}) expresses an important condition for inflation
  and can be understood from a slightly different perspective. Indeed, the constancy of
  the slow roll parameters for $V(\phi) \sim \phi^b$ 
  \begin{equation}
  \epsilon, \eta \sim \phi^{b(1-q)-2}
  \end{equation}
  immediately leads to (\ref{powerlaw}) thereby ensuring the power law inflation. The similar situation arises
  for $V(\phi) \sim \phi^{-2/q}$ in case of a tachyon field.

  \subsection{Scaling solutions}
In this subsection we shall investigate the cosmological dynamics of 
a scalar field in presence of ordinary matter. We are mainly interesting in scaling
solutions, which can exist in this model. 
  By scaling solution we mean the situation
    in which the scalar field energy density scales exactly as the power of the scale factor, $\rho_\phi
    \sim a^{-n}$, while the energy density of the perfect fluid (with equation of state
    $p_m=(\gamma-1)\rho_m$), being the dominant component, scales as
    a (possible) different power, $\rho_m \sim a^{-m}$, $m=3\gamma$.
     
     We will follow the method of Refs.\cite{Liddle} and\cite{SavTop} (see also Ref.\cite{shuntaro} on the
     related theme), where scaling solutions
     have been found in the standard and brane cosmology.
 \subsubsection{Standard scalar field}     
      Supposing that the scalar field energy density behaves as
      $\rho_\phi \sim a^{-n}$, then using the
      Klein--Gordon equation
      \begin{equation}
      \label{Kl-Gor}
      \ddot\phi+3H\dot\phi+V'(\phi)=0,
      \end{equation}
      we see that the ratio of scalar field kinetic energy density and total
      scalar field energy density remains constant
      \begin{equation}
      \label{scalrat}
      \frac{\dot{\phi}^2/2}{\rho_\phi}=\frac n6.
      \end{equation}
       
       In case of matter dominance we have
       \begin{equation}
       \label{scalfac}
       a(t) \sim t^{\frac 2{q m}}.
       \end{equation}
       Then using Eqs. (\ref{scalfac}) and (\ref{Kl-Gor}) we get
       \begin{equation}
       \label{ddphi}
       \ddot\phi=-\frac 6{q m}\frac 1t \dot\phi-\frac{dV}{d\phi}
       \end{equation}
       and equation (\ref{scalrat}) gives
       \begin{equation}
       \label{dphi}
       \dot\phi\sim t^{-\frac n{q m}}.
       \end{equation}
       Eq. (\ref{dphi}) readily integrates to yield
       \begin{equation}
       \label{phi}
       \phi=A t^{1-\frac n{q m}}.
       \end{equation}
       Substituting (\ref{phi}) into (\ref{ddphi}) and solving equation for $V(\phi)$
       we find the potentials, which allows the scaling behavior
       \begin{equation}
       \label{potent}
       V(\phi)=\frac{2(6(\beta-2)-q m\beta)}{(\beta -2)^2 q m\beta}A^{2-\beta}\phi^\beta,
       \end{equation}
       where
       \begin{equation}
       \label{power}
       \beta=\frac{2n}{n-q m}.
       \end{equation}
       For a given potential $V(\phi) \sim \phi^{\beta}$ the scalar field energy
       density scales as $a^n$, where
       \begin{equation}
       n=\frac{q \beta}{\beta-2} m.
       \label{ttt}
       \end{equation}

         For stability analysis of this solution we use new variables
         \begin{eqnarray}
         \label{var}
         \tau=\ln t \,, \qquad u(\tau)=\frac{\phi(\tau)}{\phi_0(\tau)} \,,  
         \qquad p(\tau)=u'(\tau) \,,
         \end{eqnarray}
         where $\phi_0(\tau)$ is the exact solution given by (\ref{phi}) and the prime denotes
         the derivative with respect to $\tau$.
          
Then we have the system of two first-order differential equation
\begin{eqnarray}
\label{stabsys}
u'&=&p \,, \nonumber\\
\qquad p'&=&\frac 2{\beta-2}\left(\frac 6{q m}-\frac \beta{\beta-2}\right) 
\left(u-u^{\beta-1}\right) \nonumber\\
 &-&\left(\frac{2+\beta}{2-\beta}+\frac 6{q m}\right)p \,,
\end{eqnarray}
and scaling solution corresponds to a critical point $(u,p)=(1,0)$. Linearizing (\ref{stabsys})
about this point we find the eigenvalues of these coupled equations
\begin{eqnarray}
\label{eigenval}
\lambda_{1,2}&=&\frac{\beta+2}{2(\beta-2)}-\frac 3{q m} \nonumber \\ 
&\pm &\sqrt{\left(\frac{\beta+2}{2(\beta-2)}
\frac 3{q m}\right)^2+\frac{2\beta}{\beta-2}
-\frac{12}{q m}}.
\end{eqnarray}
The condition for stability is given by the negativity of the real parts of both eigenvalues.\par
Now we consider some properties of these scaling solutions in more detail.
First of all, positivity of the potential (\ref{potent}) requires
           
           \begin{equation}
           \label{eq12}
           \frac 1\beta < \frac{6-q m}{12},
           \end{equation}
         and  the stability condition requires additionally
           \begin{equation}
           \label{nonstab}
           \frac{\beta+2}{\beta-2} > \frac 6{q m}.
           \end{equation}

We now discuss some consequences of (\ref{eq12},\ref{nonstab}). First of all,
we should point out that since $m=3\gamma$ with $\gamma$ been the equation of
state of ordinary matter, the value for $m$ is bounded in the interval $m \in
[0,6]$. Due to an additional degree of freedom, we have more complicated situation
than one discussed in~\cite{Liddle,SavTop}. To describe it in details, it would be convenient to discuss
 two cases~ $-$ for positive and negative $\beta$~$-$ separately.

                \noindent {\it The case of $\beta < 0$}
                 
                 \noindent  The stability condition (\ref{nonstab}) gives no further restrictions
                 to the condition for existence of scaling solutions
                 \begin{equation}
                 \label{nonstab-}
                 \beta < 2 \frac{6 + q m}{6 -q m}
                 \end{equation}
                 \noindent They exist for
                 \begin{equation}
                 \label{eq12-}
                 \beta < \frac{12}{6-q m}.
                 \end{equation}
From this equation one can see that for $q \le 1$
regardless the value of $m$, we always have scaling solution. The standard cosmology
and Gauss-Bonnet brane belong to this class. For $q>1$ the denominator
in (\ref{eq12-}) can be negative, restricting the range of $\beta$ suitably
for the scaling solution. A known example is the Randall-Sundrum brane
($q=2$), where the scaling solutions exist for $\beta<6/(3-m)$ \cite{SavTop}.

                 \noindent{\it The case of $\beta > 0$}
                  
                  \noindent In this case one can rewrite (\ref{eq12}) as follows
                  \begin{equation}
                  \label{eq12+}
                  \beta > \frac{12}{6 -q m}.           
                  \end{equation}
                  \noindent As the region $0<\beta<2$ is already excluded by (\ref{eq12+})
                  (since $q m >0$) we can
                   rewrite (\ref{nonstab}) as
                   \begin{equation}
                   \label{nonstab+}
                   \beta > 2 \frac{6 + q m}{6 - q m}.
                   \end{equation}
                   \noindent The equation (\ref{nonstab+}) is more restrictive.
                    
                    Thus we find  that for $q \le 1$ scaling solutions exists for
                    $\forall $$m$ if $\beta$ is large enough. On the other hand, 
if $q>1$,
                    then there exists no scaling solutions for the matter 
with $m>6/q$,
                    or, equivalently, $\gamma>2/q$. On the Randall-Sundrum brane,
                    scaling solutions with $\beta>0$ are absent if $m>3$ \cite{SavTop}.
                     
                     We summarize our results in the following table
                      
                      \begin{table}[h]
                      \begin{tabular}{|c|c|c|}
                      \hline & $q<1$ & $q>1$\\
                      \hline $\beta<0$ & exists for all $\beta$ & exists only for
                      $\beta<\frac{12}{6-q m}$\\
                      %\hline $0<\beta<2$ & \multicolumn{2}{|c|}{does not exist} \\
                      \hline $\beta>0$ & exists $\forall$ m, 
                      but $\beta > 2\frac{6+q
                      m}{6-q m}$ &
                              does not exist for $m>\frac {6}{q}$\\
                              \hline
                              \end{tabular}
                              \end{table}
For a  zone
 with negative $\beta$ where there are no scaling solutions
 the stable kinetic-term-dominated
 solution exists.
 It's explicit form for our model $H^2 \sim \rho^q$ is
 \begin{equation}
 \label{kindom}
 \phi \sim t^{1-\frac 6{q m}}.
 \end{equation}
   The eq. (\ref{ttt}) gives also the following. Consider first the case of
        $\beta <0$.
     For $q \le 1$ the numerator is always less then the denominator. It
     means
     that the scalar field energy density always drops less rapidly then the
     matter
     density. For $q >1$ it happens only if $\beta>2/(\alpha-1)$. On the
     other hand,
     if $\beta>0$ and $q \ge 1$, field energy density scales faster then
     matter.
     For $q<1$ it happens if $\beta>2/(1-q)$, i.e. for power-law
     potentials which can not support inflation.
     \subsubsection{Tachyon field}
      
      We would now  investigate the existence of tachyon field scaling solution for matter dominance
      in the general cosmological background described by $H^2 \sim \rho^q$. Assuming $\rho_\phi \sim a^{-n}$ for
      tachyon field energy density, one can obtain from using Eq.~(18)
      \begin{equation}
      {\dot\phi}^2=\frac n3 ,
      \end{equation}
      which  integrates to yield
      \begin{equation}
      \label{scsoltachphi}
      \phi=\sqrt{\frac n3}\,t.
      \end{equation}
       
       Since the matter energy density scales as the power of the scale factor $\rho_m \sim a^{-m}$,
       in case of matter dominance we have
       \begin{equation}
       \label{afromt}
       a(t)\sim t^{\frac 2{qm}}
       \end{equation}
       Substituting (\ref{scsoltachphi}) and (\ref{afromt}) into (17) and solving this equation for
       $V(\phi)$ we find the potentials, which allow the scaling behavior for tachyon field
       \begin{equation}
       V(\phi) \sim \phi^{-\beta},
       \end{equation}
       where
       \begin{equation}
       \label{betatach}
       \beta=\frac{2n}{qm}.
       \end{equation}
       As for the stability of the solution, we use the same new variables as in case of
       usual scalar field
       \begin{equation}
       \label{tachvar}
       \tau=\ln t, \qquad u(\tau)=\frac{\phi(\tau)}{\phi_0(\tau)}, \qquad p(\tau)=u'(\tau),
       \end{equation}
       where $\phi_0(\tau)$ is the exact solution given by (\ref{scsoltachphi}). Then we can get the system of two first-order differential
       equations as analogue of system~(71). And linearizing these equations about critical point
       $(u,p)=(1,0)$ corresponds to a scaling solution, we find the eigenvalues of these system
       \begin{eqnarray}
       \label{lambdatach}
       \lambda_{1,2}&=&\nonumber \frac12\: \big [\beta-\frac 6{qm}-1\\
       &\pm&\sqrt{1+6\left(\beta-\frac 6{qm}\right)+
       \left(\beta-\frac 6{qm}\right)^2}\:\big].
       \end{eqnarray}
       The condition for stability is as usual given by the negativity of the real parts of both
       eigenvalues.
        
	The condition of stability for tachyon scaling solution gives
	\begin{equation}
	\label{stabtach}
	0 < \beta < \frac 6{qm},
	\end{equation}
	which readily leads to $n <3$ or equivalently $\gamma <1$ if the tachyon field
	mimics the background (solutions with this property are often called {\it scaling solutions}
	or {\it trackers}). It should be noted that the
	condition for existence of stable scaling solution for tachyon is independent of $q$.
         Our result is in agreement with Ref.\cite{AL} which investigates the scaling solutions in case of standard GR
	 ($q=1$). In general, our findings for tachyon field are consistent with the results of Ref.\cite{SS} which
	provides a unified frame work to investigate the scaling solutions for a variety of dynamical systems.
     
\section{conclusions}
In this paper we have examined different aspects of scalar field dynamics in Gauss-Bonnet brane worlds. 
We have presented detailed investigation of tachyon inflation in GB background. Our analysis is quite general 
and deals with tachyon field dynamics at all energy scales from GR to GB regimes for polynomial type of potentials.
The information of GB correction is encoded in the functions ${\cal D}$, $\theta$ and the slow roll parameters. We find that 
the spectral index  reaches a maximum value in the intermediate region between RS and GB regimes which improves for lower values of the exponent in case of positive $p$. Our analytical results and numerical treatment of the full GB dynamics show that tachyonic inflation
with inverse square potential leads to scale invariant spectrum in the high energy GB limit which is in agreement with
the asymptotic analysis of Ref.\cite{gs}. 
The combined effect of GB term on the tensor to scalar ratio, encoded in $ {\cal D}(\chi)$,~$\theta(\chi)$ and the slow roll parameters, is such that $R$
peaks around the RS regime and exhibits a minimum in the intermediate region.
While evolving the energy scales from RS to high energy GB patch,
the background dynamics gradually changes from $H^2 \sim \rho^2$ to $H^2 \sim \rho^{2/3}$ mimicking the GR like features 
($H^2 \sim \rho$) in the intermediate region. This is an important property of GB correction which manifests in both tachyonic and standard scalar field 
dynamics.
We have shown
that the tensor to scalar ratio is generally very low in case of GB tachyonic inflation at all energy scales for
polynomial type of potentials with generic positive values of $p$. We find similar features in case inflation
is driven by inverse power law potentials with large negative powers. When $p$ is close to the scale invariant limit
, the tensor to scalar ratio becomes large in the high energy GB  regime whereas it is suppressed in the intermediate 
region making these models consistent with observation.\\ 
In section IV, we  have examined the generalized dynamics with Friedmann
equation $H^2 \sim \rho^q$ for an arbitrary $q$. This allowed us
to explain some known differences between the standard cosmology ($q=1$)
and a Randall-Sundrum brane ($q=2$) in the framework of a unified
picture as well as to obtain new results in the case of Gauss-Bonnet brane
($q=2/3)$. 
In the generalized background cosmology, we have investigated the asymptotic behavior of scalar field near cosmological singularity
and studied scaling solutions in the regime when a perfect
 fluid energy density dominates. 
In the case of ordinary scalar field, we have demonstrated that the underlying field 
dynamics exhibits distinct features depending whether $q < 1$ or 
$q > 1$ which, in particular, distinguishes the Gauss-Bonnet and Randall-Sundrum brane worlds.
For tachyon system, we have shown that the existence of stable scaling solutions $\forall$  $q$ is guaranteed if the adiabatic index of barotropic
fluid $\gamma <1$.
\bigskip
 \section{Acknowledgments}
We thank Gianluca Calcagni and Shinji Tsujikawa for their critical comments and for pointing out the reason for the mismatch
we had with their recent work in the asymptotic regime. MS acknowledges the useful discussion with N. Dadhich, T. Padmanabhan and V. Sahni. The work of AT was supported by RFBR grant 02-02-16817 and scientific school grant
2338.2003.2 of the Russian Ministry of Science and Technology.
\appendix
%%%%%%%%%%%%%%%%%%%%%%%%%%%%%%%%%%%%%%%%%%%%%%%%%%%%%%%%%%%%%%%%%%%%%%%%%%%%%%%%%%%%%%%%%%%%%%%%%%%%%%%%
\section{Various energy scales, the RS brane tension, the GB energy scale and all that}
 The GB brane world contains different energy scales discussed in \cite{DR} which we summarize here
   for the sake of completeness.
The GB term may be thought of as the lowest-order stringy
correction to the 5D Einstein-Hilbert action, with coupling
constant $\alpha>0$. In this case, $\alpha|\R^2|\ll|\R|$, so that
 \be
 \alpha \ll \ell^2\,, \label{aal}
  \ee
  where $\ell$ is the bulk curvature scale, $|\R|\sim \ell^{-2}$.
  The RS type models are recovered for $\alpha=0$.
   The 5D field equations following from the bulk action are
   \begin{eqnarray}
   \G_{ab} &=& -\Lambda_5 \g_{ab}+{\alpha\over 2}{\H}_{ab}\,,\label{afe}\\
   \H_{ab} &=& \left[\R^2-4\R_{cd}\R^{cd}+\R_{cdef}\R^{cdef}
   \right]\g_{ab} \nonumber\\&&~{}-4 \left[ \R\R_{ab}-
   2\R_{ac}\R^c{}_b\right.
   \nonumber\\&&~\left.{}-2\R_{acbd}\R^{cd}+\R_{acde}\R_b{}^{cde}
   \right]\,.\label{a5d}
   \end{eqnarray}
    
    An AdS$_5$ bulk satisfies the 5D field equations, with
    \begin{eqnarray}
    \bar{\R}_{abcd}&=& -{1\over \ell^2}\left[ \,^{(5)\!}\bar{g}_{ac}
    \,^{(5)\!}\bar{g}_{bd} - \,^{(5)\!}\bar{g}_{ad}
    \,^{(5)\!}\bar{g}_{bc} \right],\\ \bar{\G}_{ab} &=& {6\over
    \ell^2} \,^{(5)\!}\bar{g}_{ab}=-\Lambda_5 \,^{(5)\!}\bar{g}_{ab}
    +{\alpha\over 2}\bar{\H}_{ab}\,,\label{abg}\\ \bar{\H}_{ab} &=& {24
    \over \ell^4}\,^{(5)\!}\bar{g}_{ab}\,.
    \end{eqnarray}
    It follows that
    \begin{eqnarray}
    \Lambda_5 &=& -{6\over \ell^2} +{12\alpha\over
    \ell^4}\,,\label{alam}
    \\
    {1\over \ell^2} &\equiv& \mu^2={1\over 4\alpha} \left[1 - \sqrt{
    1+{4\over3} \alpha\Lambda_5} \right] \,,\label{aell}
    \end{eqnarray}
    where we choose in Eq.~(\ref{aell}) the branch with an RS limit,
    and $\mu$ is the energy scale associated with $\ell$. This reduces
    to the RS relation $1/\ell^2=-\Lambda_5/6$ when $\alpha=0$. Note
    that there is an upper limit to the GB coupling from
    Eq.~(\ref{aell}):
    \begin{equation}\label{alim}
    {\alpha} < {\ell^2 \over 4}\,,
    \end{equation}
    which in particular ensures that $\Lambda_5<0$.
     
     A Friedman-Robertson-Walker (FRW) brane in an AdS$_5$ bulk is a
     solution to the field and junction equations. The
     modified Friedman equation on the (spatially flat) brane
     is
      \be
      \kappa_5^2(\rho+\lambda) = 2\sqrt{H^2+\mu^2}\left[3-4\alpha\mu^2
      +8\alpha H^2\right]. \label{amf}
       \ee
       This may be rewritten as 
        \bea
	H^2 &=& {1\over
	4\alpha}\left[(1-4\alpha\mu^2)\cosh\left({2\chi\over3}
	\right)-1\right]\,,\label{amfe}\\
	\label{achi} \kappa_5^2(\rho+\lambda) &=&
	\left[{{2(1-4\alpha\mu^2)^3} \over {\alpha} }\right]^{1/2}
	\sinh\chi\,,
	 \eea
	 where $\chi$ is a dimensionless measure of the energy density.
	 Note that the limit in Eq.~(\ref{alim}) is necessary for $H^2$ to
	 be non-negative.
	  
	  When $\rho=0=H$ in Eq.~(\ref{amf}) we recover the expression for
	  the critical brane tension which achieves zero cosmological
	  constant on the brane,
	   \bea
	   \kappa_5^2\lambda &=& 2\mu(3-4\alpha\mu^2)\,. \label{asig'}
	    \eea
	    The effective 4D Newton constant is given by
	    \begin{equation}\label{ak4k5}
	    {\kappa_4^2  }= {\mu \over (1+4\alpha\mu^2) }\,\kappa_5^2\,.
	    \end{equation}
	    When Eq.~(\ref{aal}) holds, this implies $M_5^3\approx M_4^2/\ell$.
	     The modified Friedman equation~(\ref{amfe}), together with
	     Eq.~(\ref{achi}), shows that there is a characteristic GB energy
	     scale,
	      \be \label{am*}
	      M_{\rm GB}= \left[{{2(1-4\alpha\mu^2)^3} \over {\alpha} \kappa_5^4
	      }\right]^{1/8}\,,
	       \ee
	       such that the GB high energy regime ($\chi\gg1$) is $\rho+\lambda
	       \gg  M_{\rm GB}^4$. If we consider the GB term in the action as a
	       correction to RS gravity, then $ M_{\rm GB}$ is greater than the RS
	       energy scale $M_{\lambda}=\lambda^{1/4}$, which marks the transition
	       to RS high-energy corrections to 4D general relativity. By
	       Eq.~(\ref{asig'}), this requires $3\beta^3-12\beta^2+15\beta-2<0$
	       where $\beta \equiv 4\alpha\mu^2$. Thus (to 2 significant
	       figures),
	       \be \label{alim2}
	       M_{\lambda}<  M_{\rm GB}~\Rightarrow~\alpha\mu^2 < 0.038\,,
	        \ee
		which is consistent with Eq.~(\ref{aal}).
		 
		 Expanding Eq.~(\ref{amfe}) in $\chi$, we find three regimes for the
		 dynamical history of the brane universe:\\ the GB regime,
		  \be
		  \rho\gg  M_{\rm GB}^4~ \Rightarrow ~ H^2\approx \left[ {\kappa_5^2
		  \over 16\alpha}\, \rho \right]^{2/3}\,,\label{avhe}
		   \ee
		   the RS regime,
		    \be
		      M_{\rm GB}^4 \gg
		     \rho\gg\lambda \equiv M_{\lambda}^4 ~ \Rightarrow ~ H^2\approx
		     {\kappa_4^2 \over 6\lambda}\, \rho^{2}\,,\label{ahe}
		      \ee
		      the 4D regime,
		       \be
		       \rho\ll\lambda~ \Rightarrow ~ H^2\approx {\kappa_4^2 \over 3}\,
		       \rho\,. \label{agr}
		        \ee
			The GB regime, when the GB term dominates gravity at the highest
			energies, above the brane tension, can usefully be characterized
			as
			 \be\label{agbr}
			 H^2\gg \alpha^{-1} \gg \mu^2\,,~~ H^2 \propto \rho^{2/3}\,.
			  \ee
			  The brane energy density should be limited by the quantum gravity
			  limit, $\rho<M_5^4$, in the high-energy regime. By
			  Eq.~(\ref{avhe}),
			   \be
			   \rho<M_5^4 ~\Rightarrow~ H < \left({\pi M_5 \over 2\alpha}
			   \right)^{1/3}.
			    \ee
			    In addition, since $\rho \gg  M_{\rm GB}^4$, we have
			     \be
			     \label{aalphagreater} M_5 \gg  M_{\rm GB} ~\Rightarrow~ \alpha \gg {2
			     \over (8\pi M_5)^2}\,.
			      \ee
			      Combining these two equations leads to
			       \be\label{ahlim}
			        M_{\rm GB}^4\ll \rho < M_5^4 ~\Rightarrow~ H \ll 4\pi^{3/2}\, M_5\,.
			        \ee
				Comparing Eqs.~(\ref{aalphagreater}) and (\ref{alim2}), we also find
				that
				 \be
				 \ell \gg {1\over 8\pi M_5}\,,
				  \ee
				  The mass scales $M_5$ and $M_4$ are related 
				  \be
				  M_5^3\simeq \sqrt{\frac{4\pi}{3}} \lambda^{1/2}M_4 
				  \ee
				  Since the brane energy density is limited by quantum gravity limit
				  , the dimensionless energy scale $\chi$ can not exceed certain maximum value
				  $\chi_{max}$. Using COBE normalized value of density perturbations we found
				  $\chi_{max}\simeq 6$. As for the brane tension $\lambda$, it is typically
				  of the order of $10^{-5} M_4^4$. \\
				  Equations (A6) and (\ref{achi}) allow to relate
				  the scales $M_5$ and $M_{\rm GB}$
				  \be
				  M_5^{4} \simeq M_{\rm GB}^4 \sinh(\chi_{max})
				  \ee
The typical estimates for various scales are
\be
M_{\lambda} \simeq 10^{-5}M_4,~~M_5 \simeq 10^{-3}M_4,~~M_{\rm GB}\simeq 10^{-4}M_4
\ee
These estimates are consistent with the bounds on various scales in the problem quoted above. We once again
emphasize that we treat here GB term perturbatively such that the smooth limit to RS brane world exists.

%%%%%%%%%%%%%%%%%%%%%%%%%%%%%%%%%%%%%%%%%%%%%%%%%%%%%%%%%%%%%%%%%%%%%%%%%%%%%%%%%%%%%%%%%%%%%%%%%%%%%%%

\end{document}